\documentclass[aps,prb,reprint,longbibliography,superscriptaddress,nofootinbib,nobibnotes]{revtex4-2}

\usepackage{amsmath}
\usepackage{amssymb}
\usepackage{xcolor}
\usepackage{amsmath,amssymb,amsfonts,bm,color,graphicx}
\usepackage{appendix}
\usepackage{CJKutf8}
\usepackage{tabularx,multirow,array,diagbox}

\usepackage{bm}

\usepackage[unicode=true,colorlinks=true]{hyperref}

\usepackage[etex=true,export]{adjustbox}

\usepackage{ulem}

\newcommand{\beq}{\begin{equation}}
	\newcommand{\eeq}{\end{equation}}
\newcommand{\beqn}{\begin{eqnarray}}
	\newcommand{\eeqn}{\end{eqnarray}}

\begin{document}

\title{Full Gate-Voltage Control of a Parity-Protected Superconducting Qubit with an Altermagnetic Josephson Junction}

\author{Guo-Liang Guo}
\email{sjtu2459870@sjtu.edu.cn}
\affiliation{Tsung-Dao Lee Institute, Shanghai Jiao Tong University, Shanghai 200240, China}
\affiliation{Shanghai Research Center for Quantum Sciences, Shanghai 201315, China}

\begin{abstract}
Parity-protected superconducting qubits offer intrinsically long coherence, but many current implementations require magnetic-flux biasing, which introduces flux noise, control overhead, and limited scalability. Here we propose a parity-protected qubit based on a gate-tunable superconductor-altermagnet-superconductor Josephson junction. Altermagnets are compensated magnets with momentum-dependent spin splitting and zero net magnetization, providing spin-dependent functionality without external magnetic fields. In the proposed junction, the two spin sectors acquire opposite phase shifts, generating two Josephson channels whose interference is controlled electrically by the chemical potential. At the tuned $0$-$\pi$ transition, the first Josephson harmonic is strongly suppressed while the second harmonic dominates, yielding a double-well potential with two nearly degenerate states of opposite Cooper-pair parity. For realistic gatemon-compatible parameters, we estimate coherence times of up to tens of milliseconds while maintaining fully gate-controlled qubit operations. These results establish altermagnetic Josephson junctions as a promising route toward protected superconducting qubits with local, scalable, and all-electrical control.
\end{abstract}

\maketitle

{\it Introduction.$-$} The Josephson junction, a nonlinear circuit element arising from the coherent tunneling of Cooper pairs between two superconducting electrodes, is the central building block of superconducting quantum circuits and underpin the rapid progress of superconducting quantum computing  \cite{Koch2007,Arute2019,Kjaergaard2020,Wu2021,Acharya2023,Gao2025,Huang2026,Fischer2026,Jiang2026}. A major milestone was the introduction of the transmon qubit in 2007 \cite{Koch2007}, which substantially reduced charge-noise sensitivity by operating in the regime of large Josephson coupling-to-charging energy ratio. However, in superconducting desgin, strong protection against environmental noise typically requires delocalization in the relevant conjugate variable, whereas fast control and high-fidelity readout favor sufficient energy levels anharmonicity and addressability \cite{Siddiqi2021,Gyenis2021a}. The transmon qubit exponential suppress charge-noise sensitivity at the cost of reduced anharmonicity. This trade-off has motivated alternative architectures, including flux qubits \cite{Kjaergaard2020,Huang2020}, fluxonium qubits \cite{Manucharyan2009,Pop2014,Nguyen2019} and parity-protected qubits \cite{Kitaev2006,Brooks2013,Dempster2014,Groszkowski2018,Weiss2019,Guo2022,Guo2024}, which exploit different circuit mechanisms to suppress dominant noise channels at hardware level while maintaining a strong spectrum anharmonicity \cite{Bao2022a,Nie2026}. Recent progress in materials science offers additional degrees of freedom, such as spin and edge states , for controlling Josephson junctions. These capabilities create new platforms, such as gatemon\cite{Larsen2015,Lange2015,Casparis2018}, Topological insulator based Josephson junction \cite{Fu2009,Ren2019} to implement parity-protected qubits \cite{Larsen2020,Guo2024}. Nevertheless, such architectures, rely on external magnetic-field or magnetic flux biasing to access and control the relevant operating regime, which introduces flux noise, magnetic cross-talk, and substantial control overhead.



Recent advances in materials science provide new opportunities to engineer Josephson elements beyond conventional flux control. In particular, altermagnets (AMs) form a distinct class of compensated magnets with momentum-dependent spin splitting and vanishing net magnetization \cite{Bai2024,Song2025}. Unlike conventional antiferromagnets, in which opposite spin sublattices are typically related by translation or inversion and the bands remain spin degenerate, AMs can exhibit symmetry-protected spin-split electronic structures despite zero net moment. This spin splitting is tied to crystal-rotation symmetry and can arise even without an external magnetic field, giving AMs a unique combination of antiferromagnetic robustness and ferromagnet-like spin selectivity \cite{Li2026}. These features make AMs especially attractive for superconducting hybrid devices, where one seeks spin-dependent Josephson functionality without the detrimental orbital and Zeeman effects associated with applied magnetic fields. AMs have already been predicted to host unconventional superconducting and magnetoelectric responses \cite{Banerjee2024,Duan2025a,Chen2025,Lin2025}.

\begin{figure}[!htbp]
	\centering
	\includegraphics[width=1\columnwidth]{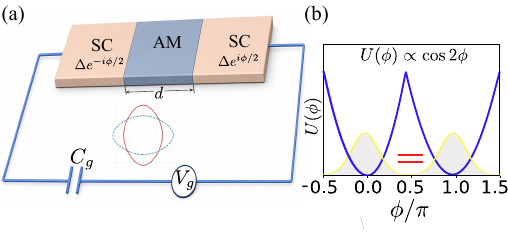}
	\caption{(a) The setup of AM Josephson junction, AM with superconductors on both ends, $\phi$ the superconducting phase difference, the red and blue curves are the Fermi surface of AM. (b) The potential shape (blue line) in $\pi$ periodic, wavefunction (light yellow) and energy levels (red lines) of AM Josephson junction at the $0-\pi$ point.
  }
\label{device-sys}
\end{figure}

In this work, we propose a parity-protected superconducting qubit based on an altermagnetic Josephson junction, operated entirely by gate-voltage control [Fig.~\ref{device-sys}(a)]. The AM-induced momentum-dependent spin splitting generates two Josephson channels with opposite phase shifts. By tuning the chemical potential, their interference drives the junction to a $0$-$\pi$ transition, where two qubit states of opposite Cooper-pair parity become nearly degenerate [Fig.~\ref{device-sys}(b)]. This electrically realizes parity protection without magnetic-flux biasing. We further show that the scheme can be implemented in a gatemon architecture and supports fully electrical initialization, control, and readout. Using realistic parameters, we estimate coherence and relaxation times in the millisecond regime.

{\it Model Hamiltonian.$-$} The proposed device is an Superconductor/AM/Superconductor (SC--AM--SC) Josephson junction in which an altermagnetic weak link is contacted by two superconducting electrodes with phase difference $\phi$. A local electrostatic gate tunes the chemical potential $\mu$ of the altermagnetic region, while a capacitive gate $V_g$ controls the offset charge $n_g$ of the effective superconducting island, as sketched in Fig.~\ref{device-sys}(a). These two voltages provide purely electrical control over both the Josephson potential and the qubit operating point, eliminating the need for magnetic flux bias or external magnetic fields.

We model the normal altermagnetic region by the tight-binding Hamiltonian
\begin{equation}
    h(\bm{k})=
    2t(2-\cos k_x-\cos k_y)s_0
    +M(\cos k_x-\cos k_y)s_z ,
\end{equation}
where $s_0$ and $s_z$ are the pauli matrix in spin space, $t$ is the hopping amplitude, and $M$ characterizes the strength of the altermagnetic spin splitting. The corresponding spin-resolved dispersions are
\begin{equation}
    E_{\pm}(\bm{k})=
    2t(2-\cos k_x-\cos k_y)
    \pm M(\cos k_x-\cos k_y),
\end{equation}
where the $\pm$ branches correspond to spin up (down) part, respectively.

In the Josephson junction geometry, this momentum-dependent splitting shifts the two spin sectors in opposite directions. As a result, the two electrons forming a Cooper pair acquire a finite center-of-mass momentum, which generates opposite phase shifts in the two Andreev channels across the junction. In the one-dimensional limit $k_y=0$ and for $M\ll t$, the momentum shift can be estimated as $q\simeq k_F M/(2t)$, with $k_F$ the Fermi momentum.  For a junction of length $d$, propagation through the altermagnetic region then produces opposite phase shifts $\pm 2qd$. Their interference modifies the current-phase relation: when $2qd \approx \pi/2$, the first Josephson harmonic is strongly suppressed, whereas the second harmonic dominant. The Josephson coupling energy is therefore driven from a conventional $\cos\phi$ form to one dominated by $\cos 2\phi$, yielding a double-well Josephson potential.


{\it Andreev levels of the Josephson junction.$-$} To obtain the Andreev spectrum, we construct the BdG Hamiltonian as 
\begin{equation}
    H_{\rm{BdG}}=[h(k)-\mu]\tau_z + \Delta\tau_x,
    \label{BdG-h}
\end{equation}
with $\tau_{x,z}$ act in Nambu space, $\mu$ the chemical potential, $\Delta$ the superconducting gap, and typically $\Delta/h=45$ GHz with Al superconductor \cite{Larsen2020}. For simplicity, we first consider the 1D limit $k_y=0$ and assuming $\Delta\ll\mu,M\ll t$, we can linearize the kinetic energy around the Fermi momenta $\pm k_f$, and split the total Hamiltonian into two decoupled part as
\begin{equation}
    \begin{aligned}
        H_{+} &=\left(\begin{array}{cc}
            v_F\left(-i\partial_x - k_F + q\right) & \Delta \\
            \Delta & -v_F\left(-i\partial_x - k_F - q\right)
        \end{array}\right) \\[10pt]
        H_{-} &=\left(\begin{array}{cc}
            v_F\left(-i\partial_x - k_F - q\right) & \Delta \\
            \Delta & -v_F\left(-i\partial_x - k_F + q\right)
        \end{array}\right)
    \end{aligned}
\end{equation}
written in the Nambu bases $(c_\uparrow,c^\dagger_\downarrow)$ and $(c_\downarrow,c^\dagger_\uparrow)$, respectively.  respectively. $k_f=\sqrt{\mu/t},v_f=2t\sqrt{\mu/t}$. $q$ denotes the altermagnet-induced momentum shift. Thus, in the low-energy description, the altermagnetic splitting enters as an electrically tunable channel momentum.

With the standard scattering matrix method \cite{Davydova2022,Mei2025}, we can calculate the eigenvalue contributed by the $H_{\pm}$ as
\begin{equation}
    E_{\pm}=\Delta\cos\left(\frac{\phi\pm2qd}{2}\right)\mp qv_f,
    \label{andreev-eq}
\end{equation}
with $d$ the junction length. The full spectrum contains both particle-hole partner eigenvalues $\pm E_{\pm}$. The zero-temperature free energy of the Josephson Junction is the summation of the occupied negative branches. A finite $q$, controlled by $\mu$ and the altermagnetic splitting $M$, shifts the two channels in opposite directions and thereby reshapes the Josephson potential.

At the qubit level, the Josephson junction acts as a tunable nonlinear element, whose potential is generally expanded in Josephson harmonics as
\begin{equation}
  V(\mu,\phi)=\sum_{n\geq 1} E_{J,n}(\mu,M)\cos(n\phi).
  \label{eq:main-potential}
\end{equation}
$E_{J,n}$ the gate-tunable fourier coefficient. Because the two spin-split channels contribute symmetrically, with identical transmission amplitudes and opposite phase shifts, the Josephson potential remains an even function of $\phi$ and contains no sine harmonics.

\begin{figure}[!htbp]
	\centering
	\includegraphics[width=1\columnwidth]{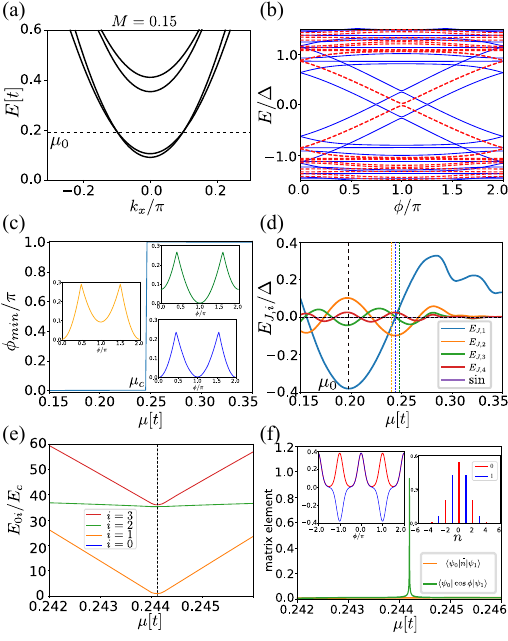}
	\caption{ (a) Band structure of the AM with periodic boundary conditions along the $x$ direction. $\mu_0$ marks zero Cooper-pair center-of-mass momentum. (b) Andreev bound-state spectrum at $\mu=\mu_0$ ($q=0$, red) and away from $\mu_0$ (finite $q$, blue). (c) The minimal point of the Josephson potential changes with $\mu$, insets shows the Josephson potential shape near the $0-\pi$ transition point, corresponding to the vertical lines of the same color in (d). (d) Leading Josephson harmonics versus $\mu$. (e) The energy spectrum $E_{0i},(i=0,1,2,3)$ changes with chemical potential near the $0$-$\pi$ transition point. (f) The transition matrix element changes with $\mu$ near the $0-\pi$ transition point, insets show the lowest two states wavefunctions in $n$ and $\phi$ space at $0-\pi$ transition point and $n_g=0$, respectively.
  }
\label{JJ-sys}
\end{figure}

Based on the analytical results, we then do the numerical calculations base on the Tight-binding (TB) model with KWANT program \cite{Groth2014}. Owing to the high transparency of the SC/AM/SC junction, only a few transverse channels in the normal region are sufficient to generate a strong Josephson coupling. Fig.~\ref{JJ-sys}(a) shows the band structure with the periodic condition in $x$-direction, open boundary in the $y$ direction. For a transverse mode $k_y=n\pi/(N_y+1)$ ($N_y$ the sites number along $y$ direction), the spin degeneracy occurs at finite momenta rather than at $k_x=0$. The corresponding chemical potential $\mu_0=4t(1-\cos k_y)$ marks the point at which the Cooper-pair center-of-mass momentum vanishes. Importantly, $\mu_0$ is only determined by the transverse quantization and is therefore set by the device geometry, independent of the altermagnetic splitting strength.

As $\mu$ is tuned away from $\mu_0$, a finite pair momentum $q$ shifts the Andreev levels [Fig.~\ref{JJ-sys}(b)] in agreement with Eq.~\eqref{andreev-eq}. Summing the negative-energy states yields the zero-temperature Josephson potential. With increasing $\mu$, the junction evolves from a $0$ junction to a $\pi$ junction, passing through a nearly degenerate double-well landscape [Fig.~\ref{JJ-sys}(c)]. Near the $0$-$\pi$ transition, the first harmonic is strongly suppressed, and the Josephson energy is dominated by a positive $\cos2\phi$ term, producing the double-well potential required for the parity-protected qubit. With the Josephson potential, we extract the Fourier coefficients in Eq.~\eqref{eq:main-potential} by Fourier transformation, shown in Fig.~\ref{JJ-sys}(d). In the short-junction limit, tuning $\mu$ mainly changes the relative phase shifts between channels, so the harmonics vary approximately periodically around $\mu_0$, $E_{J,n}\sim \cos[kn(\mu-\mu_0)]$. Consequently, near the transition, odd harmonics are strongly suppressed, while the positive second harmonic dominates.


{\it Construction of the qubit states.$-$}In the context of the Josephson potential, including the charging energy $E_c$, we can write the effective qubit Hamiltonian as \cite{Koch2007}
\begin{equation}
    H=4E_c(\hat{n}-n_g)^2+V(\mu,\phi),
    \label{qb-h}
\end{equation}
with $\hat{n}$ the Cooper pair number operator, $n_g$ the offset charge controlled by the gate voltage $V_g$, $V(\mu,\phi)$ the Josephson potential obtained from the microscopic Josephson junction calculation. In a superconducting system, we have the canonical commutation relation $[\phi,n]=i$, $\cos n\phi$ corresponds to the cotunneling of $n$ number of Cooper pairs \cite{Larsen2020}. Fig.~\ref{JJ-sys}(e) shows the lowest four eigenenergies as a function of chemical potential for finite $E_c$ at $n_g=0$ in the transmon regime. Near the electrically tuned $0-\pi$ transition point, the system exhibits two nearly degenerate states, separated from higher energy levels by a large gap (approximately $\sqrt{32E_{J,2}E_c}$ \cite{Koch2007}) due to the two degenerate double-well potential shape, shown in Fig.~\ref{JJ-sys}(c). The small residual splitting between the two lowest levels originates from interwell tunneling induced by the finite charging energy. At the transition point, the qubit splitting is first-order insensitive to variations of chemical potential, identifying an operating sweet spot. The energy difference of the lowest two states $E_{01}$ changes periodic with $n_g$, and can be approximate as \cite{Smith2020}
\begin{equation}
E_{01}\approx16E_c\sqrt{\frac{2}{\pi}}\left(\frac{2E_{J,2}}{E_c}\right)^{3/4}e^{-\sqrt{2E_{J,2}/E_c}}\cos(\pi n_g).
\label{eq-gap}
\end{equation}
This expression shows that $n_g=0$ is also a charge sweet spot. Note that, with the $\cos2\phi$ Josephson potential form, single Cooper pair tunneling is stronglu suppressed and the system can only allow even numbers of Cooper pairs tunneling, the lowest two states consver the Cooper pair parity operator $\hat{P}=e^{i\hat{n}\pi}$. Insets in Fig.~\ref{JJ-sys}(c) shows the wavefunctions of lowest two states in Cooper pair number basis, which only located at even/oddnumber sites with expectation values $\langle \hat{P}\rangle\to \pm1$ for the lowest two states. Besides, we also calculate the wavefunction of the lowest two states in the $\phi$ basis, shown in inset panel in Fig.~\ref{JJ-sys}(f), it is mainly located around the two minima of the double-well potential $\phi=0$ and $\phi=\pi$ with the expression
\begin{equation}
  |\psi_{0(1)}\rangle\approx\frac{1}{\sqrt{2}}(|\phi\approx0\rangle\pm|\phi\approx\pi\rangle),
\end{equation}
with $|\phi \approx 0\rangle$ and $|\phi\approx\pi\rangle$ refer to the state solely localized at the potential wells $\phi \approx 0$ and $\phi \approx \pi$, respectively. With the calculated wavefunctions, we further evaluate the relevant transition matrix elements near the near the $0-\pi$ transition point. Owing to the opposite Cooper-pair parity of the two lowest states, the transition matrix element $\langle\psi_0|\hat{n}|\psi\rangle_1$ approaches zero. By contrast, the phase-dependent matrix element $\langle\psi_0|\cos\hat{\phi}|\psi_1\rangle$ remains finite and deeply relies on the chemical potential, and it approaches unit at the $0-\pi$ transition point, as shown in Fig.~\ref{JJ-sys}(f).

{\it Coherent properties of the qubit.$-$} Having established the qubit states, we then analyze its coherence properties. Since the qubit is controlled electrically through the chemical potential $\mu$ and the offset charge $n_g$, the dominant decoherence channels arise from fluctuations in these two parameters. Note that the qubit is operated at a sweet spot with respect to both $\mu$ and $n_g$. We thus expand the full qubit Hamiltonian up to the second order at the sweet spot as
\begin{equation}
  H(t)=H_0+\partial_\lambda H\,\delta\lambda(t)
  +\frac{1}{2}\partial_\lambda^2H\,\delta\lambda^2(t)+\cdots ,
  \label{noise-expansion}
\end{equation}
where $H_0$ the ideal Hamiltonian at the operating point ($n_g=0$ and the electrically tuned double-well configuration near the $0$-$\pi$ transition), $\lambda$ denotes either $\mu$ or $n_g$.

\begin{figure}[!htbp]
	\centering
	\includegraphics[width=1\columnwidth]{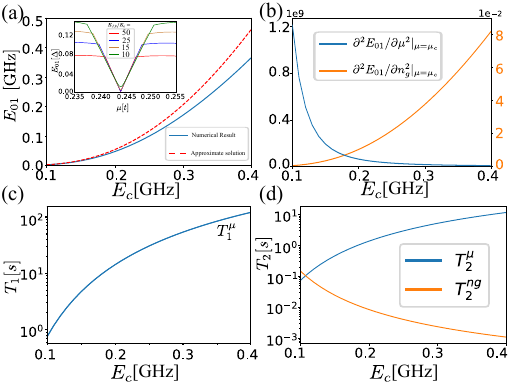}
	\caption{(a) At the $0-\pi$ transition point, $E_{01}$ change along with the variation of $E_c$, red dashed line is plotted with Eq.~\eqref{eq-gap}. Inset shows the $E_{01}$ changes with the chemical potential near the $0-\pi$ transition point with different $E_c$ (b) At the $0-\pi$ transition point, the second derivative $\partial^2E_{01}/\partial\mu^2,\partial^2E_{01}/\partial n_g^2$ as a function of $E_c$. (c) and (d) $T_1$ and $T_2$ vary with $E_c$.
  }
\label{coherent-sys}
\end{figure}

The decoherence is associated with the transitions between the qubit states, the transition rate between the initial state $|\psi_i\rangle$ to the final state $|\psi_f\rangle$ can be calculated with the Fermi's golden rule as \cite{Griffiths2018}
\begin{equation}
  \Gamma_{i\to f}=\left|\langle \psi_f|\partial_\lambda H|\psi_i\rangle\right|^2S_\lambda(\omega_{fi})
  \label{Fermi-golden}
\end{equation}
where $S_\lambda(\omega)$ is the noise power spectrum and $\omega_{fi}=E_f-E_i$. With an approximation of $1/ f$ spectrum noise, $S_\lambda(\omega) = 2\pi A_\lambda/|\omega|$ ($\omega_{ir} < \omega < \omega_{uv}$) \cite{Ithier2005,Paladino2014,Krantz2019}. Here, we estimate that $\omega_{ir}/2\pi$ = 1 Hz, $\omega_{uv}/2\pi$ = 0.4 GHz which is determined by temperature ($T_m < 20$ mK) \cite{Devoret2013,Krantz2019}, $A_{n_g}=10^{-8}e^2$ for the charge noise \cite{Koch2007}. We further assume that fluctuations in different channels are uncorrelated, so that the corresponding rates can be evaluated separately. For charge noise, we can calculate the operator $\partial_\lambda H|_{n_g=0}=8E_c\hat{n}$. Notably, the lowest two states corresponds to different Cooper pair parities, shown in Fig.~\ref{JJ-sys}(f). The transition matrix element $\langle\psi_0|\hat{n}|\psi_1\rangle$ are exactly zero due to the operator $\hat{n}$ preserves the Cooper pair parity. Moreover, the working temperature for the superconducting qubit typically $T_m\le20$ mK \cite{Devoret2013,Krantz2019}, the transition from the qubit subspace to higher excited states are exponentially suppressed by the large excitation gap. Thus, the coherent time is mainly limited by the fluctuation of the chemical potential. The corresponding noise operator can be canculated as
\begin{equation}
  \frac{\partial H}{\partial\mu}\left|_{\mu=\mu_c}=\frac{\partial E_{J,1}}{\partial\mu}\right|_{\mu=\mu_c}\cos\phi.
\end{equation}
Here, we only reatin the first order of the harmonic, as $\frac{\partial E_{J,2}}{\partial\mu}|_{\mu=\mu_c}=0$ at the sweet spot. In contrast to $\hat n$, the operator $\cos\phi$ can flip the parity of the qubit states, causing the direct transition between them. Note that, the fluctuation of chemical potential is also caused by the fluctuation of the voltage, we approximate $\mu=\kappa V$ within the lower fermi level, with $\kappa$ the lever arm. This allows us to relate the chemical-potential noise amplitude to the charge-noise amplitude through $A_\mu=\kappa^2(2e/C_g)^2A_{n_g}\approx10^{-20}$ with $\kappa\approx10^{-2}eV/V,C_g\approx3\sim5 fF$ \cite{Koch2007}. With the expression, we calculate the coherent time $T_1^{\mu}$ as a function of the charge energy $E_c$, shown in Fig.~\ref{coherent-sys}(c). With the typical transmon regime,  the resulting relaxation time is on the order of hundreds of milliseconds.

We then consider the dephasing time $T_2$, which is related to the decay of the off-diagonal term of the density matrix \cite{Koch2007}. As the qubit is worked at the sweet spot for the chemical potential $\mu$ and offset charge $n_g$, the leading dephasing contribution arises from the second-order coupling to noise. With the standard calculation, the dephasing time is given by \cite{Ithier2005,Groszkowski2018}
\begin{equation}
    T_2 = \left[D_2^2A_\lambda^2 \ln^2\left(\frac{\omega_{uv}}{\omega_{ir}}\right)+2D_2^2A_\lambda^2\ln^2\left(\frac{1}{\omega_{ir}t}\right)\right]^{-\frac{1}{2}}, 
\label{t2}
\end{equation}
where $D_2=\partial^2\omega/\partial\lambda^2$, $t$ the time during which the qubit is exposed to noise over a single cycle, and a typical conservative value is $t=10 \mu s$ \cite{Groszkowski2018}. Noted that, there exist finit gap $\delta E$ at the working point due to the finite charge energy, and it increase with $E_c$, show in Fig.~\ref{coherent-sys}(a). Thus, increasing $E_C$ enhances the curvature of the spectrum with respect to $n_g$ while reducing its curvature with respect to $\mu$, shown in inset in Fig.~\ref{coherent-sys} (a) and calculated in Fig.~\ref{coherent-sys} (b). Consequently, $T_2^\mu$ increases with $E_C$, whereas $T_2^{n_g}$ decreases [Fig.~\ref{coherent-sys} (d)]. Combining the two contributions, we estimate an overall dephasing time $T_2$ on the order of tens of milliseconds near the operating point.

\begin{figure}[!htbp]
	\centering
	\includegraphics[width=1\columnwidth]{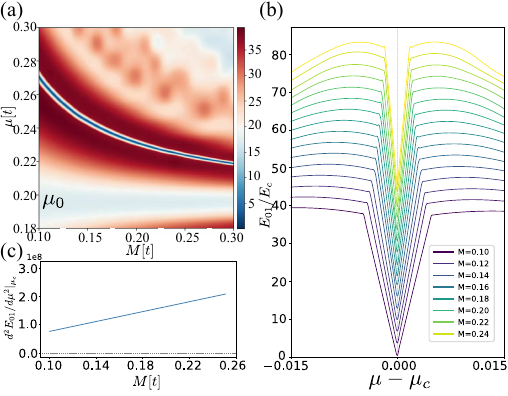}
\caption{(a) Qubit transition energy $E_{01}$ as a function of chemical potential $\mu$ and altermagnetic splitting strength $M$. (b) Waterfall plot of $E_{01}$ versus $\mu$ near the $0$-$\pi$ transition for several values of $M$. (c) Second derivative of the qubit transition energy with respect to chemical potential, $D_2^\mu=\partial^2 E_{01}/\partial\mu^2|_{\mu=\mu_c}$, evaluated at the critical chemical potential $\mu_c$ for the $0$-$\pi$ transition, as a function of $M$.}
\label{D2-sys}
\end{figure}

Finally, we examine how the qubit properties depend on the AM splitting magnitude $M$. Fig.~\ref{D2-sys}(a) shows the qubit transition energy $E_{01}$ as a function of $M$ and chemical potential $\mu$. As discussed above, the chemical potential $\mu_0$ corresponding to vanishing Cooper pair momentum $q=0$ is determined solely by the transverse quantization and is independent of $M$. By contrast, the critical chemical potential $\mu_c$ for for the electrically driven $0$-$\pi$ transition shifts to lower values as $M$ increases. Figure~\ref{D2-sys}(b) presents a waterfall plot of $E_{01}$ versus $\mu$ near the transition point for different $M$, showing that the spectral variation near $\mu_c$ becomes sharper for larger $M$. Accordingly, the second derivative of $E_{01}$ at $\mu_c$, shown in Fig.~\ref{D2-sys}(c), increases with $M$. From Eq.~\eqref{t2}, this enhanced curvature leads to stronger sensitivity to low-frequency fluctuations and hence a shorter dephasing time. Therefore, provided that the critical chemical potential remains experimentally accessible, smaller $M$ is favorable for achieving longer coherence.

{\it Operation of the qubit.$-$} Having established the qubit coherence, we now discuss its control and readout. One control scheme exploits the finite matrix element $\langle\psi_0|\cos\phi|\psi_1\rangle$ near the $0$-$\pi$ transition. Detuning the chemical potential slightly from criticality, $\mu=\mu_c+\delta\mu$, induces a first-harmonic Josephson term $\cos\phi$ that generates a transverse coupling in the qubit subspace,
\begin{equation}
	H_x=h_x(\delta\mu)\sigma_x,
\end{equation}
with $h_x(\delta\mu)$ proportional to the detuning-induced change in the first harmonic. This term drives qubit $X$ rotations. Near $\mu_c$, $E_{J,1}$ is most sensitive to $\mu$, while $\langle\psi_0|\cos\phi|\psi_1\rangle\approx 1$ [Fig.~\ref{JJ-sys}(d) and (f)], so a small $\delta\mu$ can produce a GHz-scale $h_x$ and hence nanosecond $\pi$ pulses.
A complementary control scheme follows from the wave-function structure: the qubit states are localized near $\phi=0$ and $\phi=\pi$, connected by a $\pi$ phase shift. Using $[\phi,n]=i$, this operation is generated by $e^{i\pi\hat n}$ and can be implemented by voltage driving \cite{Paolo2019}. Projected onto the qubit subspace, it realizes a Pauli-$Z$ operation. Readout can be performed using standard superconducting-circuit techniques developed for protected and transmon-like qubits \cite{Paolo2019,Smith2020,Gyenis2021}.

{\it Conclusion and discussion.$-$} We have proposed a fully gate-controlled, parity-protected superconducting qubit based on an altermagnetic Josephson junction. The essential ingredient is the momentum-dependent spin splitting in the altermagnetic region, which generates electrically tunable phase shifts in the Andreev spectrum without external magnetic fields. These phase shifts reshape the Josephson potential and can suppress the first harmonic relative to the second, thereby realizing an effective $0$-$\pi$ operating point. The gatemon architecture provides a natural platform for implementing this scheme.

The proposed qubit combines parity protection with local electrical control. Our analysis shows that relaxation and dephasing are mainly limited by chemical-potential and offset-charge noise, while the protected operating point suppresses the leading transition matrix elements. The AM can be realized with the material MnTe, CrSb \cite{Lee2024,Ding2024,Krempasky2024}. For realistic device parameters, coherence times can reach the millisecond regime, whereas voltage-pulse control enables gate operations on nanosecond timescales. Recent demonstrations of strain control in altermagnetic materials \cite{Duan2025} may offer an additional route for qubit manipulation \cite{BratlandTjernshaugen2026}. These results identify altermagnetic Josephson junctions as a promising platform for electrically controlled protected superconducting qubits.

\begin{acknowledgments}
	We acknowledge useful discussions with Xin Liu, Xun-Jiang Luo, Jiang-Hua Ying and Bei Jiang.
\end{acknowledgments}

\begin{appendix}
	
\section{Derivation of Andreev Levels in BdG Hamiltonian}

Here, we use scattering matrix to calculate the Andreev level of the system. For simplicity, we taking $k_y=0$ and consider the 1D case. In $(e \uparrow, e \downarrow, h \uparrow, h \downarrow)$ basis, the BdG Hamiltonian takes the form
\begin{equation}
H_{BdG} = 
\begin{pmatrix}
t k_x^2 s_0 + M k_x^2 s_z - \mu & \Delta \\
\Delta & -(t k_x^2 s_0 + M k_x^2 s_z - \mu)
\end{pmatrix},
\end{equation}
where $\mu$ the chemical potential, $\Delta$ the superconducting gap. In particle-hole basis, the system can be divide into two decoupled subspaces $(e \uparrow, h \downarrow), (e \downarrow, h \uparrow)$. In the limit $M < t$, we can linearize the Hamiltonian near the Fermi surface, and it becomes the form:

\begin{equation}
	\begin{aligned}
		H_{+} &=\left(\begin{array}{cc}
			v_F\left(-i\partial_x - k_F + q\right) & \Delta \\
			\Delta & -v_F\left(-i\partial_x - k_F - q\right)
		\end{array}\right) \\[10pt]
		H_{-} &=\left(\begin{array}{cc}
			v_F\left(-i\partial_x - k_F - q\right) & \Delta \\
			\Delta & -v_F\left(-i\partial_x - k_F + q\right)
		\end{array}\right)
	\end{aligned}
\end{equation}

Then, we use scattering matrix to calculate the Andreev bound states of the system. With the energy spectrum, we can get the incident electron wavefunction in the normal regime as
\begin{equation}
\psi_{in}(x) = 
\begin{pmatrix}
1 \\ 0
\end{pmatrix}
e^{i (k_F - q) x},
\end{equation}
with $k_F$ the fermi momentum, $q$ apperas due to AM. The reflected wavefunction can also be written as
\begin{equation}
\psi_{refl}(x) = b_+
\begin{pmatrix}
1 \\ 0
\end{pmatrix}
e^{-i (k_F - q) x}
+
a_+
\begin{pmatrix}
0 \\ 1
\end{pmatrix}
e^{i (k_F + q) x}
\end{equation}
The electrons can also transimit into the superconductor, and the transmitted wavefunction takes the form
\begin{equation}
\psi_{trans}(x) = c_+ 
\begin{pmatrix}
u_0^e \\ v_0^e
\end{pmatrix}
e^{i \tilde{k}_e x}
+ d_+
\begin{pmatrix}
v_0^0 \\ u_0^0
\end{pmatrix}
e^{-i \tilde{k}_h x},
\end{equation}
where $\tilde{k}_e = k_F - \frac{1}{v_F} \sqrt{(q v_F - E)^2 - \Delta^2},\tilde{k}_h = k_F + \frac{1}{v_F} \sqrt{(q v_F - E)^2 - \Delta^2}$ the momentum of the quasiparticle, $\quad
u_{0}^2 = \frac{1}{2} + \frac{\sqrt{(q v_F - E)^2 - \Delta^2}}{2 (q v_F - E)}, v_0^2 = 1 - u_0^2$ the electron (hole) part wavefunctions, respectively.

With the boundary conditions, we have the equation
\begin{equation}
\begin{aligned}
\psi_{in}(0) + \psi_{refl}(0) &= \psi_{trans}(0)\\
\psi'_{in}(0) + \psi'_{refl}(0) &= \psi'_{trans}(0)
\end{aligned}
\end{equation}

In the absence of normal reflection, the Andreev reflection coefficients can be calculated as
\begin{equation}
\begin{aligned}
a_+ &= \frac{\sqrt{(q v_F - E)^2 - \Delta^2}}{\Delta} + \frac{E - q v_F}{\Delta}, \quad\\
a_- &= \frac{\sqrt{(q v_F + E)^2 - \Delta^2}}{\Delta} + \frac{E + q v_F}{\Delta}
\end{aligned}
\end{equation}
and the scattering matrices take the form
\begin{equation}
R_- = 
\begin{pmatrix}
0 & a_- e^{i \frac{\phi}{2}} \\
a_+ e^{-i \frac{\phi}{2}} & 0
\end{pmatrix},
\quad
R_+ =
\begin{pmatrix}
0 & a_- e^{-i \frac{\phi}{2}} \\
a_+ e^{i \frac{\phi}{2}} & 0
\end{pmatrix}
\end{equation}

In the conditions without impurity scattering, the transmission matrix in normal regime can be written as
\begin{equation}
T_+ =
\begin{pmatrix}
e^{i(k_F - q) x} & 0 \\
0 & e^{i(k_F + q) x}
\end{pmatrix},
\quad
T_- =
\begin{pmatrix}
e^{-i (k_F - q) x} & 0 \\
0 & e^{-i (k_F + q) x}
\end{pmatrix}
\end{equation}

The Andreev levels can be calculated with the condition:
\begin{equation}
\det(1 - R_- T_- R_+ T_+) = 0,
\end{equation}
which gives the energy
\begin{equation}
E = \pm \left[ \Delta \cos \left(\frac{\phi + 2 q d}{2}\right) - q v_F \right]
\end{equation}
where nonzero $q$ causes interference between two Andreev levels.

\section{Numerical calculation}

\begin{figure}[!htbp]
	\centering
	\includegraphics[width=1\columnwidth]{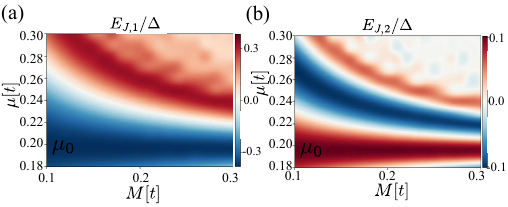}
	\caption{(a) and (b) shows the Fourier coefficients $E_{J,1}$, $E_{J,2}$ change with chemical potential and AM strength $M$.}
	\label{coef-2d}
\end{figure}

For the AM based Josephson junction, we construct a tight-binding (TB) model and perform numerical simulations using the KWANT package \cite{Groth2014}. The parameters used in the calculations are $d=50a$, $W=7a$, $\Delta=0.007t$, and $L=325a$. where $d$ the junction length, $W$ the junction width, $L$ the length of left and right superconducting leads, $\Delta$ the superconducting gap. With these parameters, the superconducting coherence length is estimated as $\xi=\hbar v_f/2\Delta\approx71a$, which gives $d\approx0.7\xi$. Typically, with the material InSb, $m^*=0.024m_e$, the electron density $n=2.7\times10^{11} cm^{-2}$ \cite{Ke2019}, the chemical potential can be estimate $\mu=\pi\hbar^2n/m^*\approx23$ meV. With typical Al superconductor, $\Delta=0.18$ meV, the coherent length can be estimated as $\xi\approx1000$ nm. Furthermore, using a typical gate lever arm for a two-dimensional electron gas, $\kappa\approx10^{-2}eV/V$ \cite{Luryi1988,Xia2009}, the gate voltage required to tune the chemical potential $V_{\mu}$ is approximately $2\sim3$ V. Fig.~\ref{coef-2d} shows the leading Fourier coefficients change with the chemical potential and AM strength. Fig.~5 in the main text is plotted with these calculated parameters.

\end{appendix}




\bibliographystyle{apsrev4-2}
\bibliography{main-ref}

\end{document}